ARTICLE

# Dose and dose-rate dependence of mutation frequency under long-term exposure

## - A new look at DDREF from WAM model -

Takahiro Wada[*], Yuichiro Manabe[a], Issei Nakamura[b], Yuichi Tsunoyama[c], Hiroo Nakajima[d], and Masako Bando[e, f]

*Department of Pure and Applied Physics, Kansai University,*

*3-3-35 Yamate-cho, Suita, Osaka 564-8680, Japan;*

[a]*Division of Sustainable Energy and Environmental Engineering, Graduate School of Engineering, Osaka University, Yamada-oka, Suita, Osaka 565-0871, Japan;*

[b]*State Key Laboratory of Polymer Physics and Chemistry, Changchun Institute of Applied Chemistry, Chinese Academy of Sciences, 5625 Renmin Street, Changchun 130022, P. R. China;*

[c]*Division of Biology, Radioisotope Research Center, Kyoto University, Sakyo-ku, Kyoto 606-8502, Japan;*

[d]*Department of Radiation Biology and Medical Genetics, Graduate School of Medicine, Osaka University, Yamada-oka, Suita, Osaka 565-0871, Japan;*

[e]*Yukawa Institute for Theoretical Physics, Kyoto University, Kitashirakawa oiwake-cho, Sakyo-ku, Kyoto 606-8502, Japan;*

[f]*Research Center for Nuclear Physics (RCNP), Osaka University, Mihogaoka, Ibaraki, Osaka 567-0047, Japan;*

[*]Corresponding author. Email: wadataka@kansai-u.ac.jp
.

**Acknowledgements**

This work was supported by Chubei Ito funds and **JSPS** KAKENHI Grant Number 15K12204 and 15K14291.

.

We investigate the dose and dose-rate dependence of the mutation frequency caused by artificial radiations with Whack-A-Mole (WAM) model which we have recently proposed. In particular, we pay special attention to the case of long-term and low dose-rate exposure. The results indicate that the dose-rate dependence is successfully described with WAM model and it may replace the so-called DDREF, the concept of which has long been adopted to take account of the difference between the high dose-rate data and the low dose-rate ones. Basic properties of WAM model are discussed emphasizing the dose-rate dependence to demonstrate how the explicit dose-rate dependence built in the model plays a key role. By adopting the parameters that are determined to fit the mega mouse experiments, biological effects of long-term exposure to extremely low dose-rate radiation are discussed. In WAM model, the effects of the long-term exposure show a saturation property, which makes a clear distinction from the LNT hypothesis which predicts a linear increase of the effects with time.

*dose rate; mutation; LNT hypothesis; DDREF; mathematical model; long-term exposure; radiation protection; radiation damage;*

# 1. Introduction

It is nowadays an urgent problem and a matter of debate how to estimate the biological effects caused by artificial irradiation. Historically, it may be traced back to the beginning of radiation genetics. The discovery of mutations in mature spermatozoa of *Drosophila* induced by artificial radiation was made by Muller in 1927 and since then, the concept of "linear no threshold (LNT)" hypothesis has long been adopted as a scientific basis of biological damage [1]. It states that the mutation caused by the ionizing radiation increases purely in proportion to the total dose, irrespectively of the dose rate. This indicated that individual ionizations within irradiated material are responsible for the mutations and the hit theory was soon formulated by Lea which explained the linear dependence of the mutation frequency upon the total dose in the low dose region [2]. The model has been adopted as a basic formula in radiation genetics. The LNT hypothesis was not only adopted by scientific communities, but it also affected the societies and governments through the 1956 BEAR I report on the risks of low doses of ionizing radiation [3]. It was summarized by the US National Academy of Sciences (NAS) Biological Effects of Atomic Radiation Committee in which the radiation-induced cancer risks were also placed within the linear hypothesis.

However, in 1958, W. L. Russell, L. B. Russell and E. D. Kelly reported that low dose-rate irradiation was less mutagenic than high dose-rate irradiation in spermatogonia and oocytes of mice [4]. They continued large-scale experiments using millions of mice to examine genetic effects of radiation, which is called mega-mouse project [5]. They observed the dose-rate effect in immature cells but not in mature spermatozoa [4]. This finding would have changed the LNT view because it indicates that there are yet other mechanisms operating in living organisms.

According to the concept of molecular and cellular responses, the general aspect of dose-response process occurring in biological objects is described in terms of stimulus-response procedure. Because this is a purely physical process, it is a reasonable consequence that the effect of a stimulus is proportional to its strength, namely to the total dose. This is quite natural if we understand what is meant by Lea's target hit theory. Note that LNT is just an approximate form for the region where the stimulus (total dose) is so weak as to neglect the higher order terms.

More importantly, we should note that the above formulation is guaranteed only for the purely physical processes without any defense system. In living organisms, however, it is commonly observed that the produced mutated cells are reduced by the competition with such defense effects. Those models which do not take account of such biological mechanism cannot be applied to the phenomena, such as mutation. They may be applicable only to the limited cases with very high dose rate experiments where the defense system is futile. Indeed, the result of mega-mouse experiments clearly indicates that the mutation frequency varies with the dose rate; the higher the dose rate is, the larger becomes the mutation frequency [5]. In order to derive human risk-estimates based on the animal experiments, it is critically important to understand the mechanism of the dose-rate effects correctly.

The recognition of the dose-rate dependence directed people to the linear-quadratic model (LQ model), which was used to estimate the difference between high and low dose-rate data in terms of the so-called DDREF (dose and dose-rate effectiveness factor) [6, 7]. An explicit form of LQ model is given in BEIR VII report for the excess effects ($E$) caused by artificial radiation with total dose $D$ as a modified version of LNT,

$$E(D) = \alpha D + \beta D^2. \quad (1)$$

This is called "linear-quadratic dose-response relationship". The quadratic term represents the contribution from two-hit processes in the hit theory. According to BEIR VII report, Equation (1) is used for the high total dose. For low total dose case, the first term is dominant and they assume that this linear dependence should be used for low dose-rate case;

$$\lim_{D \to 0} E(D) \approx \alpha D. \quad (2)$$

DDREF is used to estimate the effects of low dose-rate radiation based on the high dose-rate data. They defined the DDREF by dividing Equation (1) by Equation (2),

$$\text{DDREF} = \frac{\alpha D + \beta D^2}{\alpha D} = 1 + \frac{\beta}{\alpha} D. \quad (3)$$

Note that the above defined DDREF is obviously dependent on the total dose *D*. There is no consensus as to the numerical value as well as the range of dose and dose rate to which the DDREF should be applied[1]. Different expert groups have proposed different values for DDREF such as 1 or 2, or even 10. Moreover, this formula does not explain the fact that we obtain different results with different dose rates even for the same total dose.

**2. WAM model**

*2.1. Quick review of WAM model*

In the previous papers, we develop novel rate equations to study biological effects caused by artificial radiation exposure, accounting for the DNA damage and repair simultaneously [9, 10, 11, 12]. We call these equations Whack-A-Mole (WAM) model. The introduction of the dependence on the dose rate is critically important to take account of the defense effects which protect living organisms from mutation, such as DNA repair, apoptosis, and so on. Importantly, WAM model shows the saturation of mutation frequencies, which marks a substantial difference from existing theories based purely on the total dose.

Let us make a quick review of our model. We denote the total effect of the environment including the artificial radiation as *F*(*t*). Note that *F*(*t*) is the total effects, while *E*(*D*) counts only the excess part of the effect. For simplicity, in this paper, we take the case of a constant dose-rate, *d*, then, in WAM model, *F*(*t*) is given as a solution of the following differential equation,

$$\frac{dF(t)}{dt} = A - BF(t), \quad A = a_0 + a_1 d, B = b_0 + b_1 d. \quad (4)$$

---

[1] In its latest basic recommendations, ICRP defines low dose as 0.2 Gy or below and low dose rate as 0.1 Gy/hr or below [8].

where *A* is the source term and *B* represents the "decay" constant. In this study, *F* denotes the fraction of mutated cells, *A* is the creation rate of mutated cells and *B* is the decay rate. We treat the mutated cells which remain after repair processes, *i.e.*, the effect of DNA repair is implicitly included in the source term *A*. The decay term *B* represents the processes that reduce the number of mutated cells such as cell death. Each of *A* and *B* is expressed in terms of two components, dose-rate independent and linear ones. The former arises from the natural effect that exists even when there is no artificial irradiation (this is often called "spontaneous" term) and the latter represents the responsive effect caused by artificial irradiation. Here, we define the effective dose rate $d_{eff}$ which is the irradiation strength (dose rate) to cause the equivalent effect to the "spontaneous" term, namely, it corresponds to the stimulus of natural surroundings,

$$a_0 = a_1 d_{eff} \rightarrow A = a_1(d + d_{eff}), \quad (5)$$

In Sec. 2.3, we demonstrate that the value of $d_{eff}$ can be determined from experimental data. Note that the "spontaneous" term includes not only the effect of natural radiation, but also other factors such as miscopy of DNA in duplication process. Actually, Muller showed that the natural mutation frequency is almost $10^3$-$10^4$ times larger than the one caused by the natural radioactivity [13].

*2.2. Solution of WAM equation*

The solution of Equation (4) is easily obtained as,

$$F(t) = \frac{A}{B}\left(1 - e^{-Bt}\right) + F(0)e^{-Bt}. \quad (6)$$

This is of the same form as the one known as "growth function", which is commonly used in performing empirical fitting of plant or animal growth data between theory and experiment [14]. Characteristically, it approaches asymptotically to a certain value when *B* is positive. The asymptotic value is given by *A/B* and the time scale that WAM solution deviates considerably from the linear behavior is determined by the parameter *B*. Explicit form of the dose-rate dependence of the above time scale and of the asymptotic value are given as,

$$t_c = \frac{1}{B} = \frac{1}{b_0 + b_1 d}, \quad F(\infty) = \frac{A}{B} = \frac{a_0 + a_1 d}{b_0 + b_1 d}. \quad (7)$$

Equation (6) shows that WAM model solution describes not only the growth but also the decay of the effect. The latter occurs when the initial value $F(0)$ is much larger than the asymptotic value $A/B$, as in the case of a radiotherapy. It should be noted that, in the case of WAM model, the above parameters $A$ and $B$ which determine the behavior of $F(t)$ vary with the dose rate. Even when we have no stimulus from artificial radiation we still suffer DNA damage from natural environment which is called "spontaneous effect" and it is given as,

$$F_s = \frac{a_0}{b_0}, \quad (8)$$

Let us put $F(0) = F_s$ and define the excess effects of artificial radiation $E(t)$ as,

$$E(t) \equiv F(t) - F_S = \left(\frac{a_0 + a_1 d}{b_0 + b_1 d} - \frac{a_0}{b_0}\right)(1 - e^{-(b_0 + b_1 d)t}). \quad (9)$$

When the irradiation time $t$ is short, i.e., $t \ll t_c$, WAM model predicts linear dependence on the total dose $D$ without a threshold;

$$E(t) \approx \left(a_1 - \frac{a_0}{b_0} b_1\right) D, \quad D = d \cdot t. \quad (10)$$

Importantly, the slope does not depend on the dose rate, which is in agreement with LNT hypothesis. When the irradiation time $t$ is longer, WAM model begins to show the dose-rate dependence. The time $t$ that is necessary to achieve a total dose $D$ depends on the dose rate $d$ as $t = D/d$; the smaller the dose rate is, the larger the time becomes. WAM model has a saturation property which means $E(t)$ in Equation (9) starts to deviate from the linear dependence around $t = t_c$ and it approaches the asymptotic value,

$$E(\infty) = \left(a_1 - \frac{a_0}{b_0} b_1\right) \frac{d}{b_0 + b_1 d}. \quad (11)$$

In **Figure 1**, we show an example of WAM model results as functions of total dose $D$ for several dose rates. The parameters are given in **Table 1**, which are obtained from the mouse data [10, 11, 12]. It is seen, as the dose rate becomes smaller, that WAM model result deviates from the linear dependence at smaller dose $D$ and that the asymptotic value becomes smaller. In this way, WAM model reproduces the linear behavior which is seen in high dose-rate experiments, on one hand. It also predicts the significant deviation from the linear behavior for chronic low dose-rate irradiation, on the other hand. This remarkable character should be stressed especially for the low dose-rate case and indeed we have observed the deviation in low dose-rate data of mega mouse experiments [4, 5].

Table 1 around here

Figure 1 around here

Now we turn to the dose and dose-rate effectiveness factor (DDREF). WAM model tells us that DDREF should not be a constant; it varies with the total dose as well as the dose rate. To define an effectiveness factor that corresponds to DDREF in WAM model, we use the excess effect of the radiation of total dose $D$ with a constant dose rate $d$, $E(D, d)$, defined in Equation (9);

$$E(D,d) = \left(a_1 - \frac{a_0}{b_0}b_1\right)\frac{d}{b_0 + b_1 d}\left(1 - e^{-(b_0 + b_1 d)\frac{D}{d}}\right). \quad (12)$$

The effectiveness factor $\eta$ in WAM model is now defined as,

$$\eta \equiv \frac{E(D, d_{\text{ref}})}{E(D, d)} = \frac{1 - e^{-\frac{B_{\text{ref}} D}{d_{\text{ref}}}}}{B_{\text{ref}}/d_{\text{ref}}} \frac{B/d}{1 - e^{-\frac{BD}{d}}}, \quad (13)$$

where $B$ is given in Equation (4) as $B = b_0 + b_1 d$, $d_{\text{ref}}$ is the reference dose rate, and $B_{\text{ref}}$ is given as $B_{\text{ref}} = b_0 + b_1 d_{\text{ref}}$. Equation (13) clearly shows that the effectiveness factor $\eta$ depends on the dose rate $d$ as well as on the total dose $D$. It is to be noted here that the effectiveness factor $\eta$ is essentially independent of the production of the mutation which is proportional to $A$ in Equation (4), but should be regarded as a measure of the defense effects which is expressed by $B$. In **Table 2**, we show the effectiveness factor $\eta$ derived from the calculation of Figure 1. The reference dose rate is chosen to be $d_{\text{ref}} = 1$ Gy/hr. From Table 2, it is seen how the effectiveness factor with WAM model changes with the total dose $D$ and the dose rate $d$. It would be impractical to use a single number to represent the dose rate effect. In particular, when the dose rate is extremely low, such as 1 μGy/hr, the concept of DDREF would not be meaningful. Equation (13) shows that $\eta$ depends also on the reference dose rate $d_{\text{ref}}$. When the total dose is small, such that $B_{\text{ref}} D \ll d_{\text{ref}}$, the dependence on $d_{\text{ref}}$ can be eliminated, with a good accuracy, as

$$\eta \approx \frac{BD/d}{1 - e^{-BD/d}}. \quad (14)$$

Table 2 around here

### 2.3. Universality of WAM model and realistic parameters

In the previous publications [10, 11, 12], we have shown that WAM model rationalizes the experimental data for the relation between the mutation frequency and the artificial irradiation for five species: mouse, fruit fly, chrysanthemum, maize, and tradescantia [5, 15, 16, 17, 18, 19, 20]. Furthermore, by introducing the dimensionless "time" $\tau$ and the scaled effect function $\Phi(\tau)$,

$$\tau = \frac{t}{t_c} = (b_0 + b_1 d)t, \quad \Phi(\tau) = \frac{F(t) - F_s}{F(\infty) - F_s} = \frac{E(t)}{E(\infty)}, \quad (15)$$

we demonstrated that WAM model indeed reproduces the experimental data of various species as well as of various dose rates in a unified way [12]. Based on this universality, we here intend to treat the effects of the artificial radiation on human beings. Here, we adopt the values of the parameters ($a_0$, $a_1$, $b_0$, $b_1$) which are obtained by fitting Russell's mega-mouse experiments [10, 11, 12]. The mouse data have been usefully applied for the estimation of genetic hazards of radiation in men [21]. In such cases, it is more reliable to estimate a sort of ratio of the observable values, such as the doubling dose, the dose of radiation that induces a mutation frequency equal to the spontaneous frequency, not the values directly obtained from the data. They have been frequently utilized in the reports of radiation protection committees [7]. It is expected that the doubling dose in men is likely to be similar to that in the mouse [5].

The WAM model parameters obtained from the mouse data are listed in Table 1. From Table 1, the effective dose rate $d_{\text{eff}}$ is calculated as,

$$d_{\text{eff}} = \frac{a_0}{a_1} = \frac{3.24 \cdot 10^{-8}}{2.94 \cdot 10^{-5}} = 1.10 \cdot 10^{-3} \, [\text{Gy/hr}] = 1.10 \, [\text{mGy/hr}] \, , \quad (16)$$

and the "spontaneous" effect $F_s$ is calculated to be

$$F_s = \frac{a_0}{b_0} = \frac{3.24 \cdot 10^{-8}}{3.00 \cdot 10^{-3}} = 1.08 \cdot 10^{-5} \, . \quad (17)$$

It should be noted here that the effective dose rate $d_{\text{eff}}$ is far larger than the dose rate of natural radiation which is in the order of 0.1μGy/hr in agreement with the finding of Muller [13]. As we have already seen, both the critical time $t_c$ and the asymptotic value $F(\infty)$ depend on the dose rate. We can rewrite Equation (7) in terms of the ratio $d/d_{\text{eff}}$,

$$t_c = \frac{1}{b_0 + b_1 d} = \frac{1}{b_0 + \frac{a_0 b_1}{a_1} \frac{d}{d_{\text{eff}}}} \, ,$$

$$F(\infty) = \frac{a_0 + a_1 d}{b_0 + b_1 d} = \frac{a_0 \left(1 + \dfrac{d}{d_{\text{eff}}}\right)}{b_0 \left(1 + \dfrac{a_0 b_1}{b_0 a_1} \dfrac{d}{d_{\text{eff}}}\right)} \, . \quad (18)$$

which, for low-dose rate data, are approximately expressed for the present case,

$$t_c \approx \left(1 - \frac{a_0 b_1}{b_0 a_1} \frac{d}{d_{\text{eff}}}\right) \frac{1}{b_0} = \left(1 - 0.051 \frac{d}{d_{\text{eff}}}\right) \cdot 3.33 \cdot 10^2 \text{ [hr]},$$

$$F(\infty) \approx \left(1 + \left(1 - \frac{a_0 b_1}{b_0 a_1}\right) \frac{d}{d_{\text{eff}}}\right) \frac{a_0}{b_0} = \left(1 + 0.95 \frac{d}{d_{\text{eff}}}\right) \cdot F_s. \quad (19)$$

*2.4. WAM model and LNT hypothesis*

Now that we have seen how the low dose case differs from the high dose case, it is interesting to see how the prediction of WAM model deviates from that of a simple linear model based on LNT hypothesis. For this purpose, we show the predictions of the excess effects induced by low dose-rate radiation exposure taking account of realistic situations which we encounter in Fukushima area.

When we estimate the long-term effect of artificial radiation of low dose rate, the key quantity is the ratio $F(\infty)/F_s$ or its deviation from 1, namely, how much more effect we will get compared with the spontaneous one. From Equation (18), we obtain,

$$\frac{F(\infty)}{F_S} - 1 = \frac{E(\infty)}{F_S} = \frac{1 - \frac{a_0 b_1}{b_0 a_1}}{1 + \frac{a_0 b_1}{b_0 a_1} \frac{d}{d_{\text{eff}}}} \frac{d}{d_{\text{eff}}}$$

$$\approx \left(1 - \frac{a_0 b_1}{b_0 a_1}\left(1 + \frac{d}{d_{\text{eff}}}\right)\right) \frac{d}{d_{\text{eff}}} \approx \left(1 - \frac{a_0 b_1}{b_0 a_1}\right) \frac{d}{d_{\text{eff}}}. \quad (20)$$

Equation (20) shows that, when the dose rate of the artificial radiation $d$ is smaller than the effective dose rate $d_{\text{eff}}$, the excess effect of the artificial radiation is of the size of $d/d_{\text{eff}}$. In a realistic case of $d$ = 20 mGy/yr = 2.3 μGy/hr which is under discussion in Fukushima, $d/d_{\text{eff}}$ is estimated as 0.21%. For such a case, the effect of the artificial radiation will be too small to discriminate it from the uncertainty of the "spontaneous" effect. In this sense, the effective dose rate $d_{\text{eff}}$ is an essential quantity when we estimate the effect of artificial radiation. We have to keep in mind that we used the value of $d_{\text{eff}}$ which is derived from the mouse data. However, we expect that the variations in the effective dose rate among mammals are relatively insignificant [22]. As we mentioned earlier, it is known that if we take the ratio of observed data, such as DDREF or the doubling dose, the value is likely to be similar between men and mice [5]. The critical time which represents the time scale in WAM model is given as,

$$t_c = \frac{1}{1+\frac{a_0 b_1}{b_0 a_1}\frac{d}{d_{\text{eff}}}}\frac{1}{b_0} \approx \left(1 - \frac{a_0 b_1}{b_0 a_1}\frac{d}{d_{\text{eff}}}\right)\frac{1}{b_0} \approx \frac{1}{b_0} \quad , \quad (21)$$

It varies for different species even among mammals, because, for example, the life span is different. One may ask if the critical time $t_c$ plays any important role in the estimation of the effect of the radiation. In fact, the asymptotic value $F(\infty)$ is proportional to $t_c$, (see Equations (7)). On the other hand, the spontaneous effect $F_s$ is proportional to $1/b_0$, which is, as seen in Equation (21), essentially equal to $t_c$, for small dose rate. Thus the factor cancels in the ratio $F(\infty)/F_s$. Above discussion tells us that the difference of the critical time is not essential when we estimate the long-term effect of low dose-rate radiation.

The estimated relative effect of artificial radiation is listed in **Table 3** together with the asymptotic values for five dose rates: 0.1 μGy/hr, 1 μGy/hr, 10 μGy/hr, 100 μGy/hr, and 1 mGy/hr. In **Figure 2**, we show the time dependence of the predicted values of the mutation frequency for three dose rates: 10 μGy/hr, 100 μGy/hr, and 1 mGy/hr. For the two cases, $d$ = 0.1 μGy/hr and 1 μGy/hr, the graphs are omitted in the figure, since the excess effects are too small to be recognized in this scale. For the low dose-rate cases shown in Figure 2, the critical time does not change much from its natural value, 333 hours = 14 days, and WAM model predicts the saturation of the mutation frequency after about one month.

Figure 2 around here

Table 3 around here

On the other hand, when we estimate the same quantity with a simple linear model based on LNT hypothesis, the predictions are completely different. In **Figure 3**, we show the comparison of the predicted results of the linear model with that of WAM model. The spontaneous value $F_s$ is $1.08 \times 10^{-5}$ and we take it as the background level. The uncertainty range of the background is assumed to be about 20%, which is shown with the shaded area in Figure 3. As for the linear model, we use the following expression,

$$F_{\text{LNT}}(t) = F_s + E_{\text{LNT}}(D) = F_s + \alpha D, \quad D = d \cdot t. \quad (22)$$

We use Equation (10) to obtain the value of $\alpha$, which is $2.79 \times 10^{-5}$ [1/Gy]. Now we compare the results of the two models in term of the relative effect. From Table 2, we obtain $E(\infty)/F_s = 8.62 \times 10^{-4}$ for $d = 1$ μGy/hr, while the linear model gives $E_{\text{LNT}}(D)/F_s = 2.27 \times 10^{-2}$ after 1 year and $2.27 \times 10^{-1}$ after 10 years. For $d = 10$ μGy/hr, the linear model predicts $E_{\text{LNT}}(D)/F_s = 2.27$ after 10 years, which should be compared with $E(\infty)/F_s = 8.62 \times 10^{-3}$ with WAM model. The difference between the two models is clearly seen in Figure 3. While the predicted values of WAM model remains in the same order as the background, the predictions of the linear model increase monotonically in time and in 10 years the difference could be of two orders of magnitude, which will completely change the way how we must face the situation.

As demonstrated in the previous sections, theoretical modeling of the DNA mutation based on a single parameter (*i.e.*, the total dose) is unlikely to give in-depth insight into the mutation frequencies. Indeed, the experimental data clearly exhibit the dose-rate dependence. Thus, we suggest that one should not apply the simple linear model to low-dose radiation cases in analyzing the effects of long-term exposure.

Figure 3 around here

## 3. Conclusions

WAM model proposes a novel way to estimate the biological effects of artificial radiation. It reproduces experimental results reasonably well irrespective of the diversity of species, ranging from animals to plants. The important conclusion is that the dose rate is the fundamental quantity to measure the effects of irradiation. Because of the universality, we expect that we can apply WAM model to other species from virus to human being. One of such application is the prediction of the effects of long-term exposure to extremely low dose-rate radiation which we encounter in Fukushima.

Although the idea of LNT hypothesis is mathematically natural for small dose exposure, the linear dependence on the total dose breaks down eventually at larger total dose. The most important finding in WAM model is that the degree of the persistence of the linearity depends on the dose rate. If we have a high dose rate, the linearity persists up to a large total dose. On the other hand, when the dose rate is low, the critical total dose $D_c$ above which we obtain the saturating behavior for the excess of mutation frequency gets smaller for lower dose rate. Therefore, we should be very careful when we estimate the effect of the radiation of low dose-rate. Drawing a straight line from high dose-rate with large total dose data to make a prediction for low dose-rate case cannot be justified.

By comparing the two models, WAM model and a linear model based on LNT hypothesis, we demonstrate that the prediction based on the simple linear dependence on the total dose (LNT) is totally different from the one taking account of the dose-rate dependence (WAM). This is a serious warning that we have to take account of the dose-rate dependence when we discuss the effects of the long-term exposure to low dose-rate radiation.

Note, however, that the above discussions are based on the analyses of mutation frequency. Applicability of WAM model to health outcomes such as carcinogenesis is still an open problem which should be studied in near future.

**References**

[1] Muller HJ, Artificial Transmutation of the Gene. Science. 1927; 66: 84-87.

[2] Lea DE, Actions of Radiations on Living Cells. London: Cambridge University Press; 1947; 69-99.

[3] The Biological Effects of Atomic Radiation A Report to the Public. Washington: National Academy of Sciences; 1956.

[4] Russell WL, Russell LB, Kelly EM, Radiation Dose Rate and Mutation Frequency. Science. 1958; 128: 1546-1550.

[5] Russell WL, Kelly EM, Mutation Frequencies in Male Mice and the Estimation of Genetic Hazards of Radiation in Men. Proc. Natl. Acad. Sci. 1982; 79: 542-544.

[6] Chadwick KH, Leenhouts HP, A Molecular Theory of Cell Survival. Phys. Med. Biol. 1973; 18: 78-87.

[7] Health Risks from Exposure to Low Levels of Ionizing Radiation. Washington: The National Academic Press; 2006: 43-45.

[8] The 2007 Recommendations of the International Commission on Radiological Protection. Ann. ICRP 2007; 37.

[9] Manabe Y, Ichikawa K, Bando M, A Mathematical Model for Estimating Biological Damage Caused by Radiation. J. Phys. Soc. Jpn. 2012; 81: 104004.

[10] Manabe Y, Bando M, Comparison of Data on Mutation Frequencies of Mice Caused by Radiation with Low Dose Model. J. Phys. Soc. Jpn. 2013; 82: 094004.

[11] Manabe Y, Nakamura I, Bando Y, Reaction Rate Theory of Radiation Exposure and Scaling Hypothesis in Mutation Frequency. J. Phys. Soc. Jpn. 2014; 83: 114003.

[12] Manabe Y, Wada T, Tsunoyama Y, Nakajima H, Nakamura I, Bando M, Whack-A-Mole Model: Towards Unified Description of Biological Effect Caused by Radiation-Exposure. J. Phys. Soc. Jpn. 2015; 84: 044002.

[13] Muller HJ, Mott-Smith LM, Evidence that Natural Radioactivity is Inadequate to Explain

the Frequency of "Natural" Mutations. Proc. Natl. Acad. Sci. 1930; 16: 277-285.

[14] Richards FJ, A Flexible Growth Function for Empirical Use. J. Exp. Bot. 1959; 10: 290-300.

[15] Purdom CE, McSheehy TW, Radiation Intensity and the Induction of Mutation in Drosophila. Int. J. Rad. Biol. 1961; 3: 579-586.

[16] Spencer WP, Stern C, Experiments to Test the Validity of the Linear Drosophila at Low Dosage. Genetics. 1948; 33: 43-74.

[17] Yamaguchi H, Shimizu A, Degi K, Morishita T, Effect of Dose and Dose Rate of Gamma Ray Irradiation on Mutation Induction and Nuclear DNA Content in Chrysanthemum. Breed. Sci. 2008; 58: 331-335.

[18] Mabuchi T, Matsumura S, Dose Rate Dependence of Mutation Rates from γ-irradiated Pollen Grains of Maize. Jpn. J. Genet. 1964; 39: 131-135.

[19] Sparrow AH, Underbrink AG, Rossi HH, Mutations Induced in Tradescantia by Small Doses of X-rays and Neutrons: Analysis of Dose-Response Curves. Science. 1972; 176: 916-918.

[20] Nauman CH, Underbrink AG, Sparrow AH, Influence of Radiation Dose Rate on Somatic Mutation Induction in Tradescantia Stamen Hairs. Radiat. Res. 1975; 62: 79-96.

[21] Neel JV, Changing Perspectives on the Genetic Doubling Dose of Ionizing Radiation for Humans, Mice, and Drosophila. Teratology. 1999; 59: 216-221.

[22] Sankaranarayanan K, Chakraborty R, Ionizing Radiation and Genetic Risks XI. The Doubling Dose Estimates from the Mid-1950s to the Present and the Conceptual Change to the Use of Human Data on Spontaneous Mutation Rates and Mouse Data on Induced Mutation Rates for Doubling Dose Calculations. Mutar. Res. 2000; 453: 107-127.